# DeepConv-DTI: Prediction of drug-target interactions via deep learning with convolution on protein sequences

Ingoo Lee[¶], Jongsoo Keum[¶], and Hojung Nam[*]

School of Electrical Engineering and Computer Science, Gwangju Institute of Science and Technology, 123 Cheomdangwagi-ro, Buk-ku, Gwangju, 61005, Republic of Korea

[¶]Equally contributing authors

[*]Corresponding author

E-mail: hjnam@gist.ac.kr

**Abstract**

Identification of drug-target interactions (DTIs) plays a key role in drug discovery. The high cost and labor-intensive nature of *in vitro* and *in vivo* experiments have highlighted the importance of *in silico*-based DTI prediction approaches. In several computational models, conventional protein descriptors are shown to be not informative enough to predict accurate DTIs. Thus, in this study, we employ a convolutional neural network (CNN) on raw protein sequences to capture local residue patterns participating in DTIs. With CNN on protein sequences, our model performs better than previous protein descriptor-based models. In addition, our model performs better than the previous deep learning model for massive prediction of DTIs. By examining the pooled convolution results, we found that our model can detect binding sites of proteins for DTIs. In conclusion, our prediction model for detecting local residue patterns of target proteins successfully enriches the protein features of a raw protein sequence, yielding better prediction results than previous approaches.

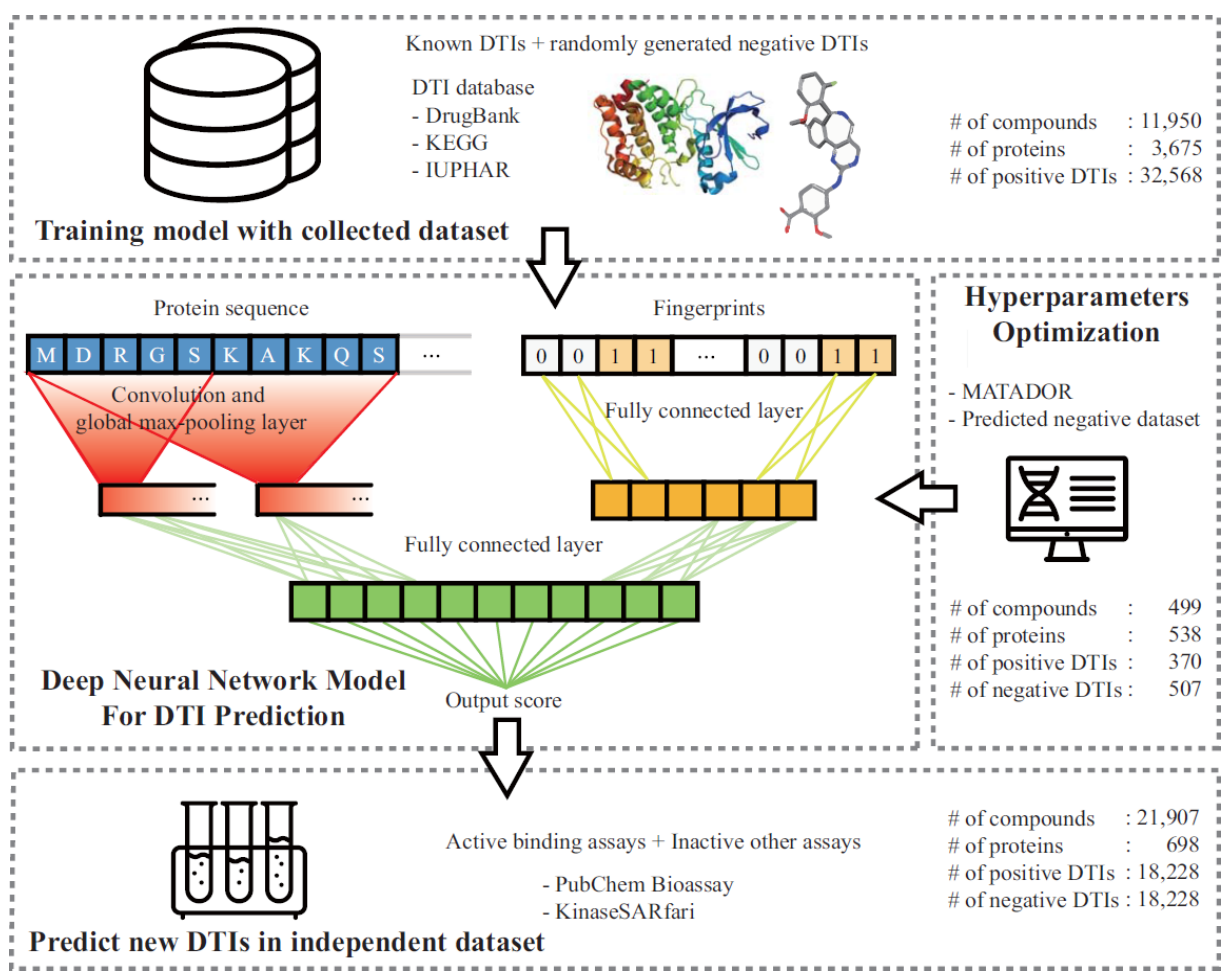

**Fig 1 . Overview of our model.** First, we collected training DTI dataset from various databases (DrugBank, KEGG, IUPHAR). Second, we constructed the neural network model using convolution, which is able to capture local residue patterns that can help the DTIs. Third, we optimized the hyperparameters with an external validation dataset that we constructed. Finally, we predicted DTIs from bioassays (independent test dataset) and evaluated the performance of our model. The number (#) of each compounds, proteins and DTIs are summarized in each step.

## Introduction

The identification of drug-target interactions (DTIs) plays a key role in the early stage of drug discovery. Thus, drug developers screen for compounds that interact with specified targets with biological activities of interest. However, the identification of DTIs in large-scale chemical or biological experiments usually takes 2~3 years of experiments, with high associated costs [1] Therefore, with the accumulation of drugs, targets, and interaction data, various computational methods have been developed for the prediction of possible DTIs to aid in drug discovery.

Among computational approaches, several similarity-based methods were initially studied, in which it was assumed that drugs bind to proteins similar to known targets and vice versa. One of the early methods is that of *Yamanashi et al.,* which used a kernel regression method to use information on known drug interactions as the input to identify new DTIs, combining a chemical space and genomic spaces into a pharmacological space [2]. To overcome the requirement of the bipartite model for massive computational power, *Beakley et al.* developed the local bipartite model, which trains the interaction model locally but not globally. In addition to substantially reducing the computational complexity, this model exhibited higher performance than the previous model [3]. As another approach to DTI prediction models, matrix factorization methods are recruited to predict DTIs, which approximate multiplying two latent matrices representing the compound and target protein to an interaction matrix and similarity score matrix [4, 5]. In this work, regularized matrix factorization methods successfully learn the manifold lying under DTIs, giving the highest performance among previous DTI prediction methods. However, similarity-based methods are not commonly used at present to predict DTIs, as researchers have found that similarity-based methods work well for DTIs within specific protein classes but not for other classes [6]. In addition, some proteins do not show strong sequence similarity with proteins sharing an identical interacting compound [7].

Thus, feature-based models that predict DTI features of drugs and targets have been studied [8-10]. For the feature-based DTI prediction models, a fingerprint is the most commonly used descriptor of the substructure of the drug [11]. With the drug fingerprint, the drug is transformed into a binary vector whose index value represents the existence of the substructure in the drug. For proteins, the composition, transition, and distribution (CTD) descriptors are conventionally used as computational representations [12]. Unfortunately, feature-based models that use the protein descriptor and drug fingerprint showed worse performance than the previous conventional quantitative structure-activity relationship (QSAR) models [8]. To improve the performance of the feature-based models, many approaches have been developed, such as the use of interactome networks [13] and minwise hashing [14]. Although various protein and chemical descriptors have been introduced, feature-based models do not show sufficiently good predictive performance [15]. For conventional machine learning models, features must be built to be readable by modeling from original raw forms, such as SMILES and amino acid sequences. During transformation, rich information, such as local residue patterns or relationships, is lost. In addition, it is hard to recover lost information using traditional machine learning models.

In recent years, many deep learning approaches have recently been developed and recruited for omic data processing [16] as well as drug discovery [17], and these approaches seem to be able to overcome limitations. For example, *Wen et al.* used the deep belief network (DBN) [18], with features such as the composition of amino acids, dipeptides, and tripeptides for proteins, and extended-connectivity fingerprint (ECFP) [19] for drugs [6]. They also discussed how deep-learning-based latent representations, which are nonlinear combinations of original features, can overcome the limitations of traditional descriptors by showing the performance in each layer. In another study [20], the researchers employed stacked autoencoder (SAE) with sparsity constraints to abstract original features into a latent representation with a small dimension. With latent representation, they trained a support vector machine (SVM), which performed better than previous methods, including feature- and similarity-based methods. In another study by *Tian et al.* [21], domain binary vectors were employed to represent the existence of domains used to describe proteins.

One way to reduce the loss of information of features is to process raw sequences and simplified molecular-input line entry system (SMILES) as their forms. In a paper by *Öztürk et al.*, DeepDTA was used to represent raw sequences and SMILES as one-hot vectors or labels [22]. With a convolutional neural network (CNN), they extracted local residue patterns to predict the binding affinity between drugs and targets. As a result, their model exhibited better performance on a kinase family bioassay dataset [23] than the previous model, kronRLS [24]. In their model, the Smith-Waterman similarity (SW) [25] score performed better than CNN on protein sequences, while CNN performed better than SIMCOMP on SMILES [26]. However, their CNN method had a bias for the maximum length of proteins because they stacked the CNN to yield local residue patterns. We can still recruit CNN to extract local residue patterns for DTI without losing generality and rich information regarding protein residues and without a bias for the maximum length of proteins.

To overcome the aforementioned problems, here, we introduce a deep learning model that predicts massive-scale DTIs using raw protein sequences not only for various target protein classes but also for diverse protein lengths. The overall pipeline of our model is depicted in Fig 1. We adopted convolution filters on the entire sequence of a protein to capture local residue patterns, which are the main protein residues participating in DTIs. By pooling the maximum results of CNN, we can determine how given protein sequences match local residue patterns participating in DTIs. Using these data as input variables of higher layers, our model constructs abstracts and organizes features for proteins. Finally, our model concatenates protein features with drug features, which come from fingerprints through the fully connected layer and predict the probability of DTIs via higher fully connected layers. We trained our model with large-scale DTI information integrated from various DTI databases, such as DrugBank, International Union of Basic and Clinical Pharmacology (IUPHAR), and Kyoto Encyclopedia of Genes and Genomes (KEGG), and we optimized the model with DTIs from MATADOR and negative interactions predicted from *Liu et al.* [27]. With the optimized model, we predicted DTIs from bioassays such as PubChem BioAssays and KinaseSARfari to estimate the performance of our model. As a result, our model exhibits better performance than a previous model that uses CTD and SW scores.

**Results and Discussion**

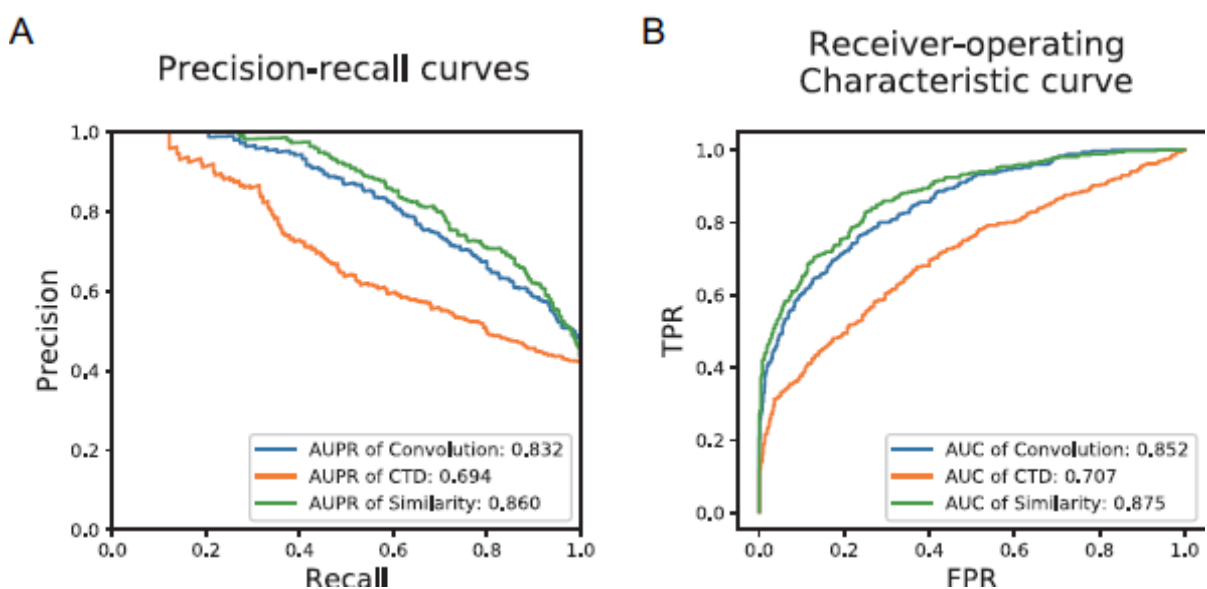

**Fig 2. Performance curves for optimized models of protein descriptors.** The AUPR and AUC of the convolution, CTD, and similarity descriptors are shown in panels A and B, respectively.

**Performances of the validation dataset and selected hyperparameters**

As a normal step of hyperparameter setting, we first tuned the learning rate of the weight update to 0.0001. After the learning rate was fixed, we benchmarked the sizes and number of windows, hidden layers of the drug features, and the concatenating layers with area under precision-recall (AUPR). Finally, we selected the hyperparameters of the modelwith the unseen validation dataset, yielding an AUPR of 0.832, as shown in Fig 2. The AUPR value of our model was less than the AUPR of the similarity descriptor; however, that does not mean that our method has lower prediction performance than the similarity method because the size of the validation dataset is too small to evaluate the general performance. In the same manner, we built and optimized models that use other protein descriptors with the same activation function, learning rate, and decay rate.

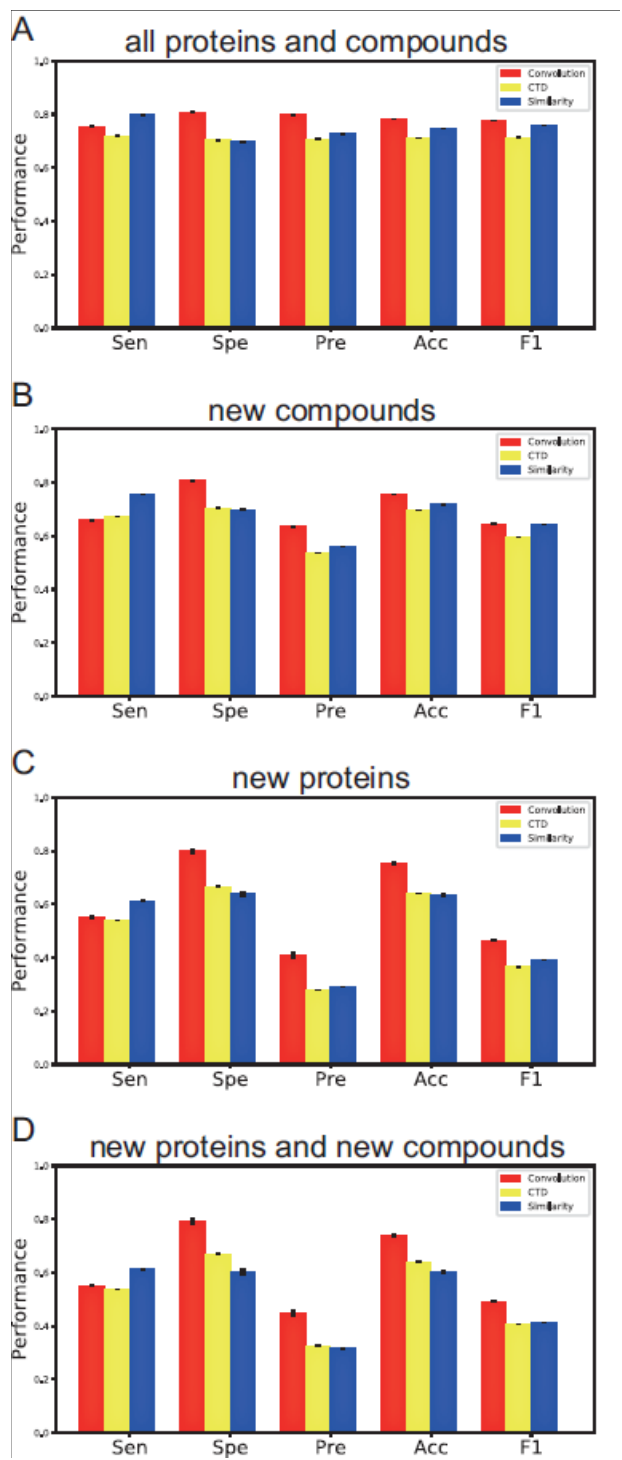

**Fig 3. Performance measures (Sen, Spe, Pre, Acc, F1) for all of the independent datasets of the PubChem dataset.** (A) All queried PubChem datasets. (B) PubChem dataset whose compounds are not in the training dataset. (C) PubChem dataset whose targets are not in the training dataset. (D) PubChem dataset whose compounds and targets are not in the training dataset. Our convolution model shows better performances for all datasets in terms of accuracy and F1 score.

**Comparison of performance with other protein descriptors**

After the hyperparameters were tuned, we compared the performance based on the independent test datasets with the different protein descriptors, the CTD descriptor (which is usually used in the conventional chemo-genomic model) [12], normalized SW score [25], and our convolution method. The results showed that our model exhibited better performance than the other protein descriptors for all the datasets as shown in Fig 3. With the threshold selected by equal error rate (EER) [28], our model performed equally well with both the PubChem and KinaseSARfari datasets, indicating that our model has general application power. Our convolution method gave the highest accuracy score and F1 score for the PubChem dataset (Fig 3A) [29] and its subsets (Fig 3B-D) and a slightly lower F1 score for the KinaseSARfari dataset [30]. The CTD descriptor gave the lowest score for any dataset and any metric, which implies that CTD is less informative and less enriched than the other descriptors. Here, we also observed that the model performance using a similarity descriptor for the KinaseSARfari dataset was similar to that of the proposed model. We can interpret this result as the similarity descriptor acts as an informative feature as a local residue pattern at the domain level, not the whole protein complex.

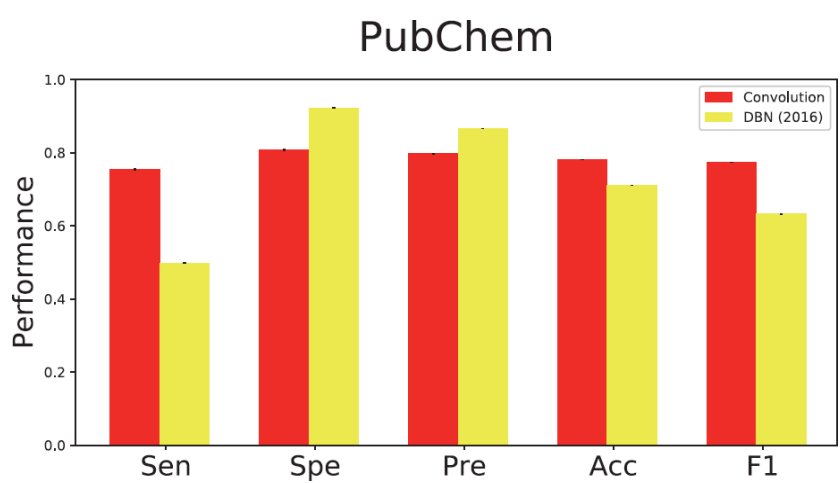

**Fig 4. Comparison of performances between our model and that of *Wen et al.***

**Performance comparison with a previous model**

In addition to the comparison between convolution in our model and other protein descriptors, we compared our model and a previous model. The previous model selected for comparison was built by *Wen et al.* [6]. Other studies have found it difficult to compare performance because their purposes and datasets are different from those of our model. We used their optimized shape hyperparameters and descriptors to represent the DTI, but we used a lower fine-tuning learning rate because the model cannot learn with a high fine-tuning learning rate of 0.1. We decreased the fine-tuning learning rate to 0.01 and executed fine-tuning until the validation error converged. After the pretraining and fine-tuning stages, we evaluated the PubChem dataset and compared the result with our model. In the comparison, our model performed better than the previous model, as shown in Fig 4. The previous model was built with deep belief network (DBN) [18], which is a stack of Restricted Boltzmann machine.

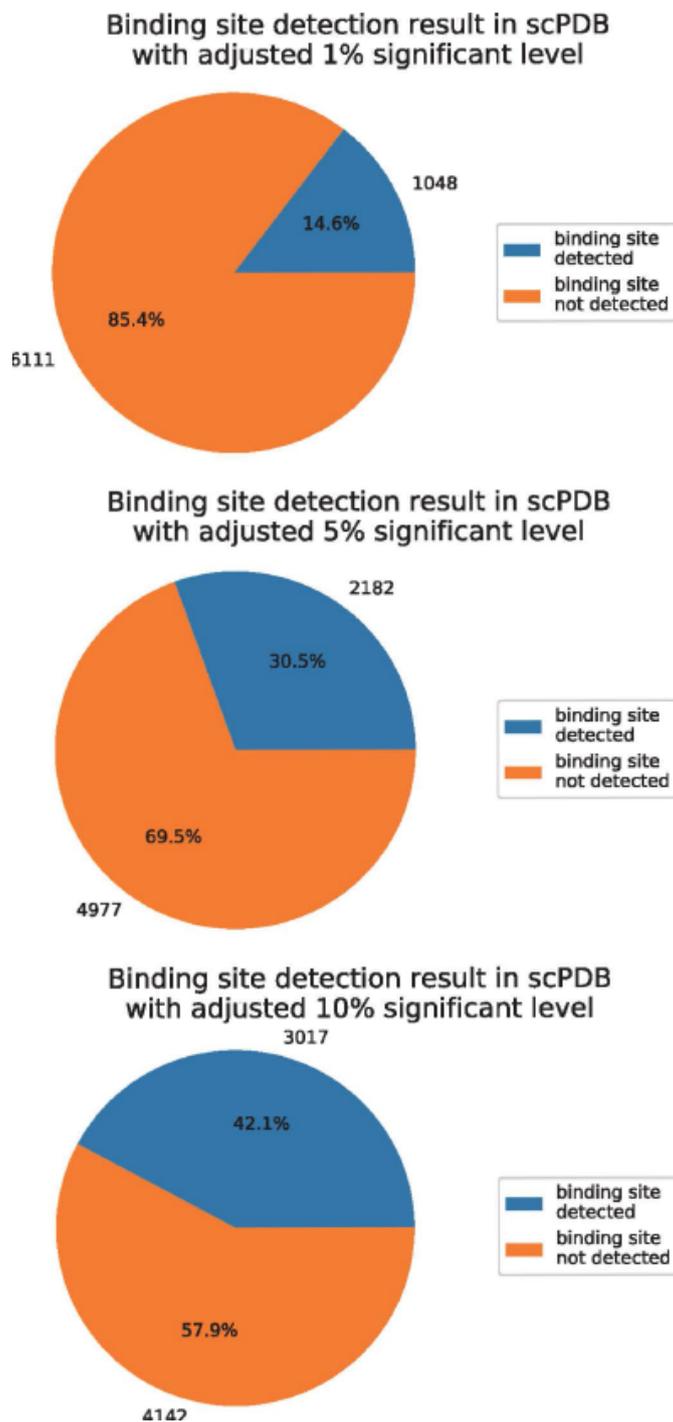

**Fig 5**. **Statistical test for binding region detection.** We executed a right-tailed *t*-test for the number of covering binding sites from the convolution results with a null distribution, which was constructed from the randomly generated convolution results in the sc-PDB database consisting of 7179. Because each sc-PDB test has many windows, we selected the most significant p-value adjusted by the Benjamini-Hochberg procedure and examined whether they were significant at levels of 1%, 5% and 10%. As a result, 6%, 15% and 22.5% of sc-PDB entries were significantly enriched, respectively.

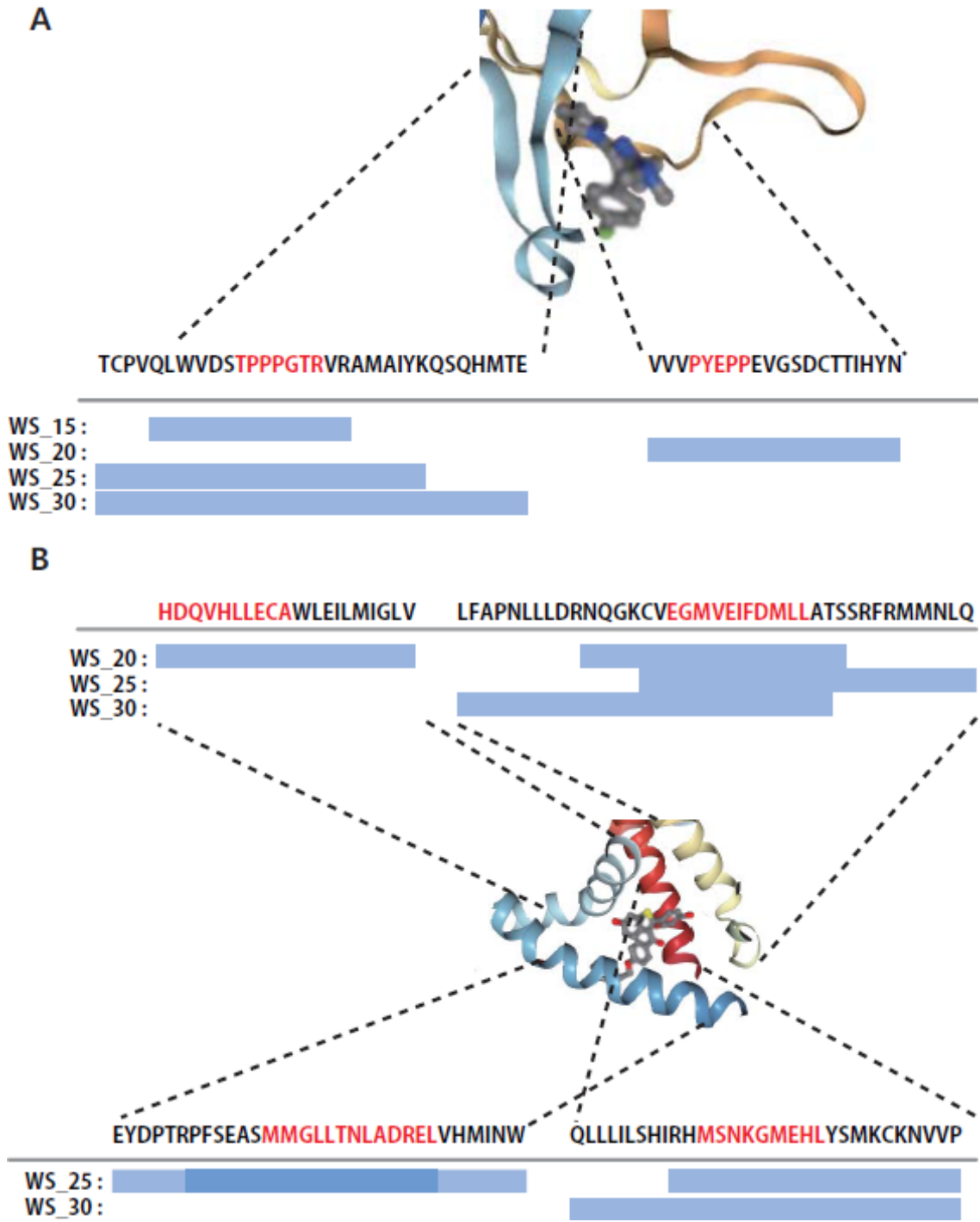

**Fig 6. Visualization of binding sites.** (A) Binding sites of the cellular tumor antigen protein (P04637, P53_HUMAN). (B) Binding sites of the Estrogen receptor protein (P03372, ESR1_HUMAN). The blurred ligand can differ by sc-PDB entry. Binding sites from sc-PDB entries are colored red. Pooled convolution results are shown as transparent blue boxes that exactly cover the binding residues.

**Detection of the binding site by CNN on the sequence**

Because we pooled the maximum convolution results by each filter for each window, the pooled results could highlight regions of matches with local residue patterns. Although we cannot measure exactly how those values affect the DTI prediction results, the pooled maximum convolution result will affect the prediction performance by going through higher fully connected layers. Therefore, if our model is capable of capturing local residue patterns, it would give high values to important regions of protein, such as actual binding sites. Examining and validating the convolution results from the intermediate layer showed that our model could capture local residue patterns that participate in DTIs. The sc-PDB database provides atom-level descriptions of proteins, ligands, and binding sites from complex structures [31]. By parsing binding site annotations, we can query binding sites between protein domains and pharmacological ligands for 7179 entries of Vertebrata. From the queried binding sites and pooled maximum convolution results, we statistically test our assumption that the pooled maximum convolution results cover the important regions, including binding sites. Each window has 128 pooled convolution results, which shows bias in covering some regions. Thus, we randomly generated 128 convolution results 10,000 times for each sc-PDB entry and counted how many of those random results covered each amino acid in binding sites, which constructed normal distributions. For each normal distribution constructed by the randomly generated convolution results, considered as a null hypothesis, we executed a right-tailed $t$-test with the number from the convolution results of our model for each window. Because we did not know which window detects the binding site, we took the most significant p-value (minimum p-value adjusted by the Benjamini-Hochberg procedure [32]).. We summarize the results of binding site detection from the most significant p-value among windows by significant level cutoff in Fig 5. In addition, we examined scPDB entries with the most significant p-values for diverse window sizes. As expected, the convolution results gave high values when binding sites were included. For example, cellular tumor antigen p53 (P04637, P53_HUMAN) has many sc-PDB entries with many pharmacological ligands. In Fig 6A, annotated binding sites from many sc-PDB entries with P53_HUMAN are colored red for each sc-PDB entry. For the annotated region, some pooled results from the convolution layers exactly cover the binding sites with high ranks among filters in a window, which must affect the prediction scores. In addition, these convolution results cover more complicated sites. For the protein Estrogen receptor (P03372, ESR1_HUMAN), which is a nuclear hormone receptor, 4 binding sites are covered with convolution results from the model with high rank among filters in windows, as shown in Fig 6B. Although we only studied binding sites, we assume that our method could enrich other important regions.

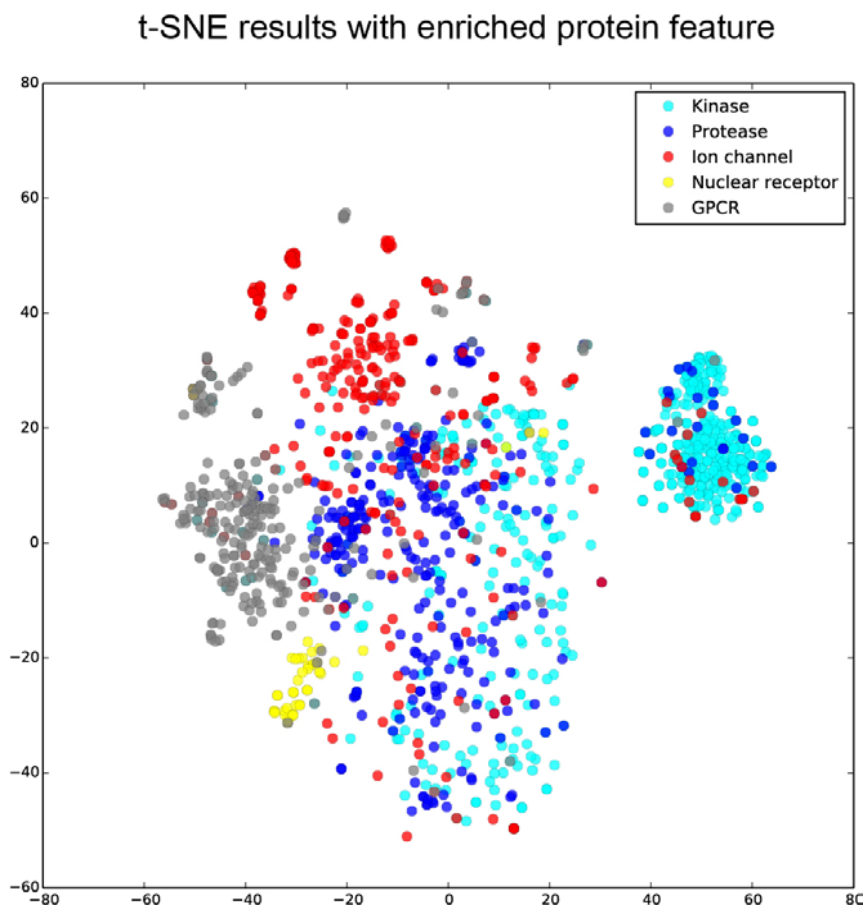

**Fig 7. t-SNE results with protein features from the fully connected layer beyond the global max-pooling layer of the convolution result.** Each color corresponds to a protein class. The visualization result shows that our model is able to roughly discriminate protein classes by capturing local residue patterns.

**t-SNE visualization of proteins**

From the results shown in Fig 6, we can confirm that our model can capture the local residue patterns of proteins that participate in DTIs. Thus, to examine further characteristics of the captured protein local residue patterns, we visualized the protein features from the fully connected layer after the global max-pooling of convolution results. We visualized 1,527 proteins used in the training dataset categorized in various protein classes. Specifically, we visualized 257 GPCRs, 44 nuclear receptors, 304 ion channel receptors, 604 kinases, and 318 protease. For the visualization, we conducted t-distributed stochastic neighbor embedding (t-SNE) for dimension reduction and visualization [33]. t-SNE can map high-dimensional features to low-dimensional ones, such as 2-dimensional features, minimizing information loss during dimension reduction. Surprisingly, although our model is not intended to identify protein classes, it can roughly discriminate protein classes from the intermediate protein layer, as shown in Fig 7.

**Materials and Methods**

**Building dataset**

To build the training dataset, we obtained known DTIs from three databases: DrugBank [34], KEGG [35], and IUPHAR [36]. To remove duplicate DTIs among the three databases, we unified the identifiers of the compounds and the proteins. For the drugs, we standardized the identifiers of the compounds in the DrugBank and KEGG databases with the InChI descriptor. For the proteins, we unified the identifiers of the proteins as UniProtKB/Swiss-Prot accessions [37]. Among the collected DTIs, we selectively removed proteins of Prokaryota and single-cell Eukaryota, retaining only proteins of Vertebrata. Finally, 11,950 compounds, 3,675 proteins, and 32,568 DTIs were obtained in total. Because all collected DTIs are regarded as positive samples for training and negative DTIs are not defined in the databases above, a random negative DTI dataset is inevitably generated. To reduce bias from the random generation of negative DTIs, we built ten sets of negative DTIs exclusively from the positive dataset.

To optimize our model with the most adequate hyperparameters, we constructed an external validation dataset that had not seen DTIs in the training phase. We collected positive DTIs from the MATADOR database [38], including 'DIRECT' protein annotations, and all the DTIs observed in the training dataset were excluded. To build a credible negative dataset, we obtained negative DTIs via the method of *Liu et al*. This method selects candidate negative DTIs with low similarity to known positive DTIs [27]. From the obtained negative dataset, we balanced the negative dataset with the positive dataset, using a negative score (>0.95). As a result, 370 positive DTIs and 507 negative DTIs were queried for the external validation set.

For the evaluation of our model, we built two independent test datasets from the PubChem BioAssay database [29] (Wang, et al., 2017) and ChEMBL KinaseSARfari [30]; these datasets consisted of results from experimental assays. To obtain positive DTIs from PubChem, we collected 'Active' DTIs from the assays with the dissociation constant ($K_d <$ 10 μ*m*)[39]. Because we sought to predict whether a drug binds to a protein, among the many types of assays (Potency, $IC_{50}$, $AC_{50}$, $EC_{50}$, $K_d$, $K_i$), the evaluation of the dissociation constant ($K_d$) was the most appropriate assay for obtaining positive samples. For the negative samples, we took the samples annotated as 'Inactive' from the other assay types. Because there were too many negative samples in the PubChem bioassays, we first collected only negative samples whose drug or target was included in the positive samples of the PubChem bioassays. Second, we selected as many random negative samples as positive DTIs from PubChem BioAssay. As a result, 18,228 positive and negative samples were built with 21,907 drugs and 698 proteins. For the performance evaluation, we created three subsets of the PubChem bioassay independent dataset for humans, which consisted of only new compounds, new proteins, and new DTIs. We also collected samples from KinaseSARfari. KinaseSARfari consists of assays involving a compound that binds to a kinase domain. To obtain positive samples from KinaseSARfari, we considered each assay result with a dissociation constant of as positive [39], which is sufficiently small to be considered positive.

In contrast to the PubChem bioassays, the number of negative samples was similar to the number of positive samples in KinaseSARfari; therefore, we did not sample the negative samples. We collected 3,835 positive samples and 5,520 negative samples with 3,379 compounds and 389 proteins. We confirmed that the training and the validation datasets were not biased toward a specific protein class.

**Drug feature representation**

In our model, we used the raw protein sequence as the input for the protein but did not use the raw SMILES string as the input for the drug. For the drug, we used the Morgan/Circular drug fingerprint, which analyzes molecules as a graph and retrieves substructures of molecular structures from subgraphs of the whole molecular graph [19]. Specifically, we used RDKit [40] to yield a Morgan/Circular fingerprint with a radius of 2 from a raw SMILES string. Finally, each drug can be represented as a binary vector with a length of 2048, whose indices indicate the existence of specific substructures.

**Selection of hyperparameters**

In our deep learning models, hyperparameters, such as the learning rate and window sizes that affect performance, are tuned during cross-validation. However, the hyperparameters should not be determined based on the performance of the subset of the training dataset because the negative datasets are randomly sampled. With the external validation dataset, we first determined the learning rate because a model with a high learning rate is unable to learn a pattern. After the learning rate was selected, we selected regularization parameters such as dropout ratio. Finally, we employed a grid-search method for optimization of the other hyperparameters. We identified hyperparameters that exhibited the best AUPR, which is an appropriate performance evaluation metric for the accuracy of classifying the positive sample. The other descriptors to compare with our methods are numerical vectors, which do not have locality. Therefore, we put fully connected layers on the protein descriptors. We also employed a grid-search strategy while sustaining hyperparameters not related to model shape. When the AUPR is measured, the optimal threshold can be given by the EER [28].

$$\text{EER} = \min_{\theta}(|1 - \text{recall}| - \gamma(1 - \text{precision})|)$$

where θ is the classification threshold and γ is the constant determining the cost ratio for misclassification from precision and recall, which is set at 2 in our model.

**Evaluation of performances**

To measure the prediction performance of our deep neural model based on the independent test dataset after the classification threshold was fixed, we obtained the following performance metrics: sensitivity (Sen), specificity (Spe), precision (Pre), accuracy (Acc), and the F1 measure (F1).

**Author Contributions**

IL, JK, and HN conceived of the study. IL and JK implemented the study. IL drafted the manuscript. HN revised and edited the manuscript. All authors have read and approved the final manuscript.